\documentclass[journal=jacsat,manuscript=article]{achemso}

\usepackage{chemformula} % Formula subscripts using \ch{}
\usepackage[T1]{fontenc} % Use modern font encodings

\author{Yali Jia $^{1}$}
\author{Zhaohua Tian$^{1}$}
\author{Qi Liu$^{1,2}$}
\author{Zhengyang Mou$^{1}$}
\author{Zihan Mo$^{1}$}
\author{Yu Tian$^{1,2}$}
\author{Qihuang Gong$^{1,2,3,4,5}$}
\author{Ying Gu$^{1,2,3,4,5}$}
 \email{ygu@pku.edu.cn}

\affiliation{$^1$State Key Laboratory for Mesoscopic Physics, Department of Physics, Peking University, Beijing 100871, China\\
$^2$Frontiers Science Center for Nano-optoelectronics $\&$  Collaborative Innovation Center of Quantum Matter $\&$ Beijing Academy of Quantum Information Sciences, Peking University, Beijing 100871, China\\
$^3$Collaborative Innovation Center of Extreme Optics, Shanxi University, Taiyuan, Shanxi 030006, China\\
$^4$Peking University Yangtze Delta Institute of Optoelectronics, Nantong 226010, China\\
$^5$Hefei National Laboratory, Hefei 230088, China}
\title{Cascade enhancement and efficient collection of single photon emission under topological protection}

\abbreviations{IR,NMR,UV}

\begin{document}

%%%%%%%%%%%%%%%%%%%%%%%%%%%%%%%%%%%%%%%%%%%%%%%%%%%%%%%%%%%%%%%%%%%%%
%% The "tocentry" environment can be used to create an entry for the
%% graphical table of contents. It is given here as some journals
%% require that it is printed as part of the abstract page. It will
%% be automatically moved as appropriate.
%%%%%%%%%%%%%%%%%%%%%%%%%%%%%%%%%%%%%%%%%%%%%%%%%%%%%%%%%%%%%%%%%%%%%
%\begin{tocentry}
%
%\begin{figure*}[htbp]
%\includegraphics[width=1\textwidth]{TOC}% Here is how to import EPS art
%
%\end{figure*}

%\end{tocentry}

%\begin{tocentry}
%\includegraphics[width=6cm]{TOC}% Here is how to import EPS art
%\end{tocentry}

%Some journals require a graphical entry for the Table of Contents.
%This should be laid out ``print ready'' so that the sizing of the
%text is correct.
%
%Inside the \texttt{tocentry} environment, the font used is Helvetica
%8\,pt, as required by \emph{Journal of the American Chemical
%Society}.
%
%The surrounding frame is 9\,cm by 3.5\,cm, which is the maximum
%permitted for  \emph{Journal of the American Chemical Society}
%graphical table of content entries. The box will not resize if the
%
%content is too big: instead it will overflow the edge of the box.
%
%This box and the associated title will always be printed on a
%separate page at the end of the document.

%%%%%%%%%%%%%%%%%%%%%%%%%%%%%%%%%%%%%%%%%%%%%%%%%%%%%%%%%%%%%%%%%%%%%
%% The abstract environment will automatically gobble the contents
%% if an abstract is not used by the target journal.
%%%%%%%%%%%%%%%%%%%%%%%%%%%%%%%%%%%%%%%%%%%%%%%%%%%%%%%%%%%%%%%%%%%%%
\begin{abstract}
High emission rate, high collection efficiency, and immunity to the defects are the requirements of implementing on-chip single photon sources. 
Here, we theoretically demonstrate that both cascade enhancement and high collection efficiency of emitted photons from single emitter can be achieved simultaneously in topological photonic crystal containing a resonant dielectric nanodisk. 
The nanodisk excited by a magnetic emitter can be regarded as a large equivalent magnetic dipole. The near-field overlapping between this equivalent magnetic dipole and edge state enables to achieve a cascade enhancement of single photon emission  with Purcell factor exceeding $4\times10{^3}$.
These emitted photons are guided into edge states with collection efficiency of more than 90\%, which is also corresponding to quantum yield due to topological anti-scattering and the absence of absorption. 
The proposed mechanism under topological protection has potential applications in on-chip light-matter interaction, quantum light sources, and nanolasers. 
\end{abstract}

%%%%%%%%%%%%%%%%%%%%%%%%%%%%%%%%%%%%%%%%%%%%%%%%%%%%%%%%%%%%%%%%%%%%%
%% Start the main part of the manuscript here.
%%%%%%%%%%%%%%%%%%%%%%%%%%%%%%%%%%%%%%%%%%%%%%%%%%%%%%%%%%%%%%%%%%%%%
\section{Introduction}
Single photon sources in micro- and nanoscale are essential for on-chip quantum information processing. 
Enhancement of spontaneous emission through the optical modes in cavities, i.e., the Purcell effect, is one of the basic principles for realizing single photon sources~\cite{f1}. 
High photon emission rate, high collective efficiency, high quantum yield, and robustness to defects and perturbations are essential elements for the practical single photon sources~\cite{f2,f3}. 
In order to meet the requirements above, various micro-nano photonic structures supporting diverse optical modes have been proposed. 
Both photonic crystal (PC) cavities by confining light in defects~\cite{f4,f5} and PC waveguides with slow light near the photonic band edge~\cite{f6,f7} enhance the spontaneous emission. But owing to large mode volume, the maximum enhancement of the emission rate is only several tens of $\gamma_{0}$~\cite{N1}, where $\gamma_{0}$ is the spontaneous emission rate in vacuum. 
Although  there is a high collection efficiency of emitted photons in PC waveguide~\cite{f8,f9,f10},  low emission rate makes it difficult  to be used in on-chip photonic devices.
The hotspots induced by  surface plasmon polaritons (SPPs) in metallic nanoparticles can achieve high emission rate of more than $10^4\gamma_{0}$~\cite{f11,f12,f13,f14,f15}. While, its radiation part accounts for a small proportion due to the large absorption. There is higher extraction efficiency in  the SPP waveguide~\cite{f16,f17,f18}, but the large absorption loss leads to low quantum yield, hindering its application.
In addition, dielectric structures overcoming the loss problem and supporting multipole resonances can increase the enhancement of electric and magnetic emission up to several hundreds of $\gamma_{0}$, but the emitted photons are difficult to be collected and utilized~\cite{f19,f20,f21,f22,f23}. 

In order to achieve both high emission rate and collection efficiency, hybrid micro- and nanostructures with abundant near-fields are considered~\cite{S1,S2,S3,S4}.
Because SPPs structure has a smaller mode volume, to obtain superior properties, it becomes a good candidate to be combined with other micro-nano structures.
Firstly, owing to the ultra-small mode volume at the nanoscale gap, the gap SPPs structure provides large spontaneous emission enhancement~\cite{S5,S6}. By combining it with low-loss optical fiber, ultra-high photon emission rate and one-dimensional nanoscale photon propagation are simultaneously obtained~\cite{S7}.
Secondly, jointing  SPPs structure with a high-quality factor whispering gallery modes not only has strong emission enhancement ~\cite{S8,S9}  but also has a mixed-mode linewidth tunability~\cite{S10}.
Thirdly, in a hybrid structure of PC and SPPs structure, the local field can be greatly enhanced ~\cite{S11}, so that a large Purcell factor can be acheived~\cite{S12, S13}.
%Then in the photonic crystal, by creating the nanocavity with strong localized fields and high helicity, the chiral Purcell enhancement was achieved ~\cite{S13}.
 %In addition, photonic crystal cavities with collection waveguide can also collect photons~\cite{S14}.
From above hybrid structures, one can see that,  due to the existence of SPP, a large emission rate can be obtained, but the loss of metal materials makes the quantum yield low. 
Moreover, the properties of these micro-nano cavities are affected by defects and disturbances during the manufacturing process, which further limits their application in high-quality single photon sources.

\begin{figure*}[htbp]
\includegraphics[width=1\textwidth]{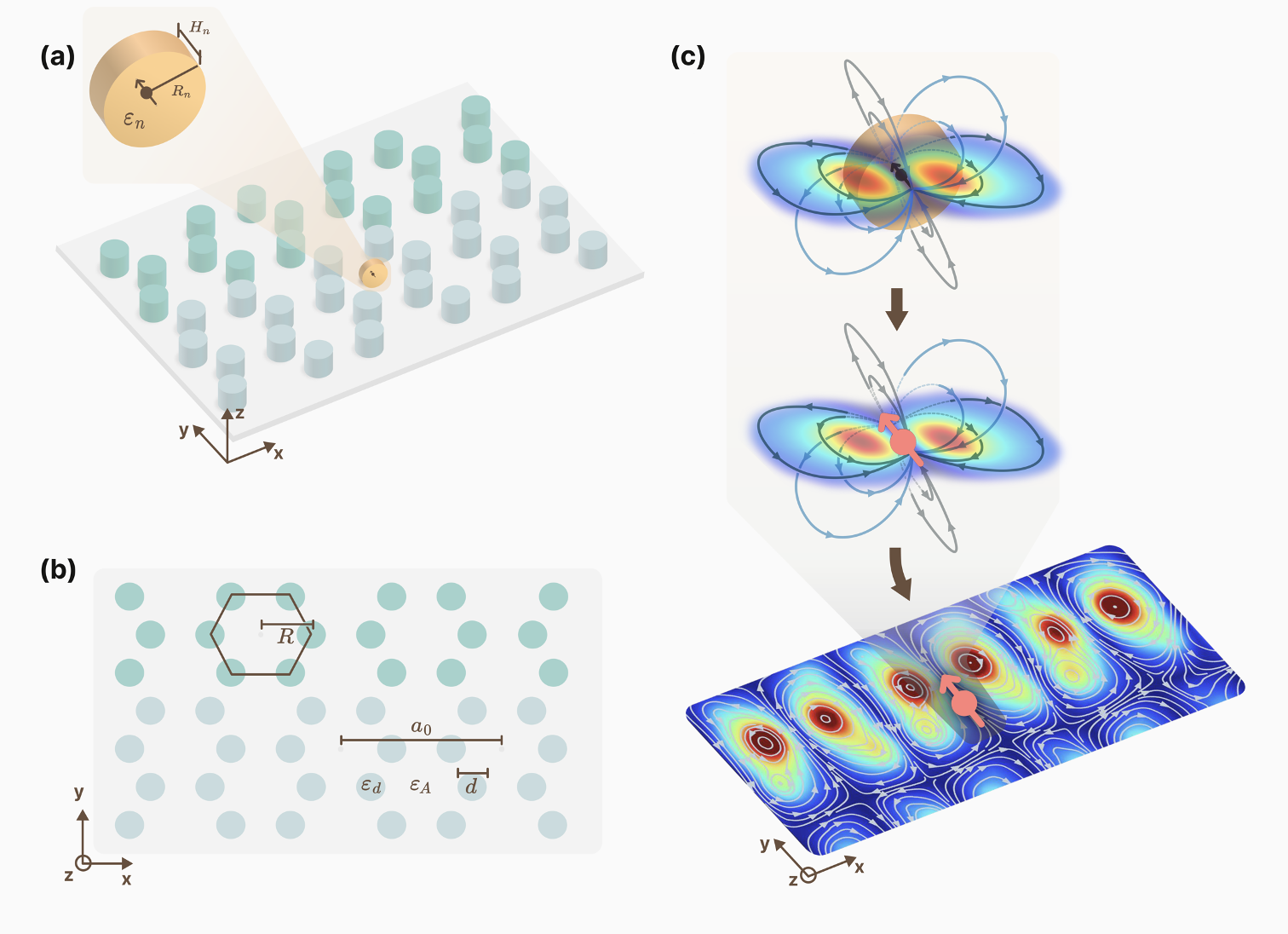}% Here is how to import EPS art
\caption{ Scheme of cascade Purcell enhancement under topological protection. (a) Schematic diagram of the topological PC-resonant nanodisk hybrid structure. A resonant nanodisk with radius $R_n$, height $H_n$, and dielectric constant $\varepsilon_n$ are inserted at the splices of PCs with different topological properties. The quantum emitter is located in the center of the nanodisk. (b) Schematic diagram of the honeycomb PC, where $R$ is the length of the hexagonal side and $d$ is the diameter of the dielectric cylinder. $\varepsilon_d$ and $\varepsilon_A$ are the dielectric constant of the dielectric cylinder and the surrounding environment, respectively.
(c) The magnetic dipole resonance of the nanodisk excited by the magnetic quantum emitter (brown arrow) can be equivalent to a large magnetic dipole (pink arrow) interacting with the edge state. The streamlines around the nanodisk and the equivalent magnetic dipole are magnetic field lines. The bottom is the electric field distribution of the edge state, and the white streamlines are the magnetic field lines.
}\label{fig1}
\end{figure*}

The unique properties of topological photonics, including topological robustness and anti-scattering,  bring new opportunities for the development of single photon sources~\cite{T1,T2,T3,T4,T5}. 
Barik et al. first proposed the chiral coupling between quantum emitters and edge states in topological PC~\cite{T6}. 
After that, Purcell enhancement is demonstrated topological corner states~\cite{T7,T8}, topological slow light in valley PCs~\cite{T9,T10}, and topological Su-Schrieffer-Heeger  cavity modes ~\cite{T11}. 
Unfortunately, their Purcell factor is not very high, almost the same order as that in photonic crystals. 
Recently, non-scattering large Purcell enhancement is obtained in the topological photonic structure containing a plasmonic nanoantenna~\cite{T12}, but still suffered from low quantum yield due to an intrinsic metallic loss. 
So, that both high photon emission rates and high quantum yield simultaneously exist under topological protection, remains an issue.

\begin{figure*}[htbp]
\includegraphics[width=1\textwidth]{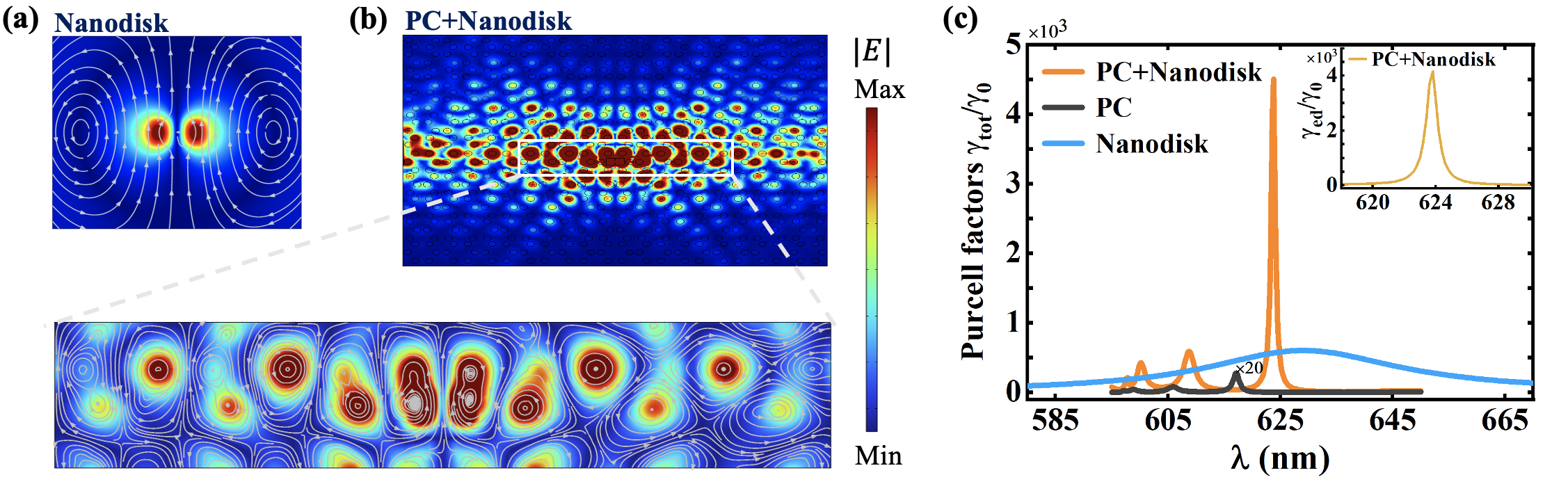}% Here is how to import EPS art
\caption{ Cascade enhancement of emission and efficient collection of photons in topological hybrid structures.
(a) The electric field in the xy plane of the magnetic dipole resonance in a bare nanodisk, where the white streamlines indicate the direction of the magnetic field.
(b) The electric field of the edge state in the xy plane of the topological hybrid structure (top part). The region where the edge state is plotted is 6$a_0\times 1.2a_0$, where the white streamlines indicate the direction of the magnetic field (bottom part).
(c) The total Purcell factors $\gamma_{tot}/\gamma_{0}$ in the bare nanodisk, bare topological PC, and the hybrid structure. The inset is the Purcell factor $\gamma_{ed}/\gamma_{0}$ along the edge state propagation in the hybrid structure.
Purcell cascade enhancement can be achieved in the hybrid structure, with a cascade enhancement factor of up to 0.53. The efficiency of collecting photons along the edge states can reach 93\%. }\label{fig2}
\end{figure*}

In the following, by combining robust edge states in topological PC with magnetic dipole resonance in low-loss dielectric nanodisk, taking advantage of both, we theoretically achieve cascade enhancement  and high quantum yield  of single photon emission under topological protection [Fig. 1(a)].
The resonance of nanodisk excited by a magnetic emitter can be regarded as a large equivalent magnetic dipole  [Fig. 1(c)]. 
Superior to metallic nanoparticles, this nanodisk has a large Purcell factor but without any nonradiative part. 
Fortunately, there is  large near-field overlapping between this equivalent magnetic dipole and topological edge state appearing at the interface of two PCs.
As a result, we  achieve a cascade enhancement of single photon emission with Purcell factor exceeding $4\times10{^3}$.
Also owing to the near-field overlapping and topological robustness, the photons scattered around the nanodisk are guided into edge states with the collection efficiency  of more than 90\%.
Since there is no absorption loss and the edge state has the property of anti-scattering,   in principle, the values of collection efficiency and quantum yield should be equal.
The proposed mechanism for bright single-photon emission and  high quantum yield under topological protection opens up the opportunities for practical applications of on-chip light-matter interaction, quantum light sources, and nanolasers.

\section{Results and discussion}

%\section{Physical considerations for model building}\label{Sec2}
Edge states with robust and anti-scattering properties are supported in topological PC. However, the Purcell enhancement is only several tens of  $\gamma_{0}$ in the bare topological PC due to the relatively extended mode volume of the edge states. This limits its application in efficient quantum information processing~\cite{T9,T10}.
Low-loss dielectric nanostructures support magnetic dipole resonance with a small mode volume~\cite{M1}. However, photons scattered around are difficult to collect and utilize. 
Therefore, here we combine the advantages of both and design a hybrid structure of topological PC containing a resonant nanodisk [Fig. 1(a)].  Cascade enhancement of emission and efficient collection of emitted photons are achieved simultaneously. 

In the topological hybrid structure, the quantum emitter is placed in the near field region of the nanodisk, the total decay rate
 can be divided into two parts $\gamma_{tot}=\gamma_{ed}+\gamma_{sc}$, where $\gamma_{ed}$ is the part that decays to the edge state and $\gamma_{sc}$ is the part that radiates into the free space. For a nanodisk in free space, all photons are scattered into free space, i.e. $\gamma_{tot}=\gamma_{sc}$.
In the following, we use COMSOL Multiphysics software to calculate the Purcell factor (defined as $PF=\gamma/\gamma_{0}$) of each part.  The calculation details are given in the  Supplemental Material~\cite{SM}. %

Consider the hybrid structure of topological PC containing a resonant nanodisk [Fig. 1(a)], where the honeycomb topological PC with $C_{6}$ symmetry was proposed by Xiao Hu's group~\cite{M2}. The primitive cell of the PC is shown in Fig. 1(b), which is a hexagonal honeycomb composed of six cylinders. Here $a_0$ is the lattice constant and $R$ is the distance from the center of the primitive cell to the center of the dielectric cylinder. In the cell, the diameter of the dielectric cylinder is $d=65$ nm and the dielectric constant is $\varepsilon_d=11.7$, the rest is air with the dielectric constant $\varepsilon_A=1$.
When the lattice constant $a_0$ is fixed at 319 nm, $R$ is adjusted.
When $a_0/R=3.3$, it is a topologically trivial PC (the upper PC in Fig. 1(b)). When $a_0/R=2.7$, it is a topologically non-trivial PC (the lower PC in in Fig. 1(b)). By splicing two photonic crystals with different topological properties together, edge states appear at about $\lambda=595\sim650$ nm.
In the edge state channel, a dielectric nanodisk with a dielectric constant of $\varepsilon_n=16$, radius of $R_n=75$ nm, and height of $H_n=89$ nm is inserted, with its axial direction being the y-axis. The nanodisk supports a magnetic dipole resonance with $\lambda=629.2$ nm, within the spectral range of edge state.
The y-polarized magnetic quantum emitter is located at the center of the nanodisk in the topological hybrid structure. We used COMSOL Multiphysics software to simulate the spectral properties of edge states in topological PCs and the magnetic dipole resonance of nanodisk~\cite{SM}.

%\section{Cascade Enhancement}\label{Sec2}
We now investigate  the cascade enhancement of magnetic emission in hybrid structure.
As shown in Fig. 2, when a quantum emitter is placed at the center of a nanodisk, the maximum magnetic Purcell factor 
$\gamma_{tot}/\gamma_{0}$ can reach 4504. While  in bare topological PC, the maximum Purcell factor is only 14. When there is only a dielectric nanodisk, the Purcell factor at the resonance is 604. Therefore, much stronger emission enhancement can be achieved in the hybrid structure than that of the bare topological PC and the bare dielectric nanodisk. 
To quantitatively describe the enhancement effect in the hybrid structure, we define the cascade enhancement factor $\delta =\frac{PF_{{Hybrid}} }{PF_{ PC} \ast PF_{Nanodisk}}$. Here the cascade enhancement factor can reach $\delta =0.53$. 

The mechanism of achieving magnetic emission cascade enhancement  is described as follows. The magnetic dipole resonance supported by a dielectric nanodisk [Fig. 2(a)] can be effectively excited by a magnetic emitter, which can be equivalent to a large magnetic dipole [Fig. 1(c)].
This large magnetic dipole and the edge state have magnetic fields with similar rotational characteristics and the near-field overlapping [Fig. 2(b)], so the magnetic emission cascade enhancement is achieved.
The magnetic Purcell factor in topological hybrid structure is much larger than that in the dielectric structure proposed in Ref.\cite{M3}.

\begin{table*}[htbp]
\caption{The Purell factor $\gamma_{tot}/\gamma_{0}$, cascade enhancement factor $\delta $, collection efficiency $\beta$, and near-field overlapping degree $\eta $   when the nanodisk is rotated around the $x$-axis by an angle $\theta$ (inset in TABLE).
The parameters of the nanodisck are $\varepsilon_n=16$, $R_n$=75 nm, $H_n=89$ nm.
As $\theta$ increases, $\gamma_{tot}/\gamma_{0}$, $\delta$, and $\beta$ all decrease due to the decrease of the near-field overlapping $\eta$ between the dipole resonance of the nanodisk and the edge state.}\label{fig4}
\includegraphics[width=1\textwidth]{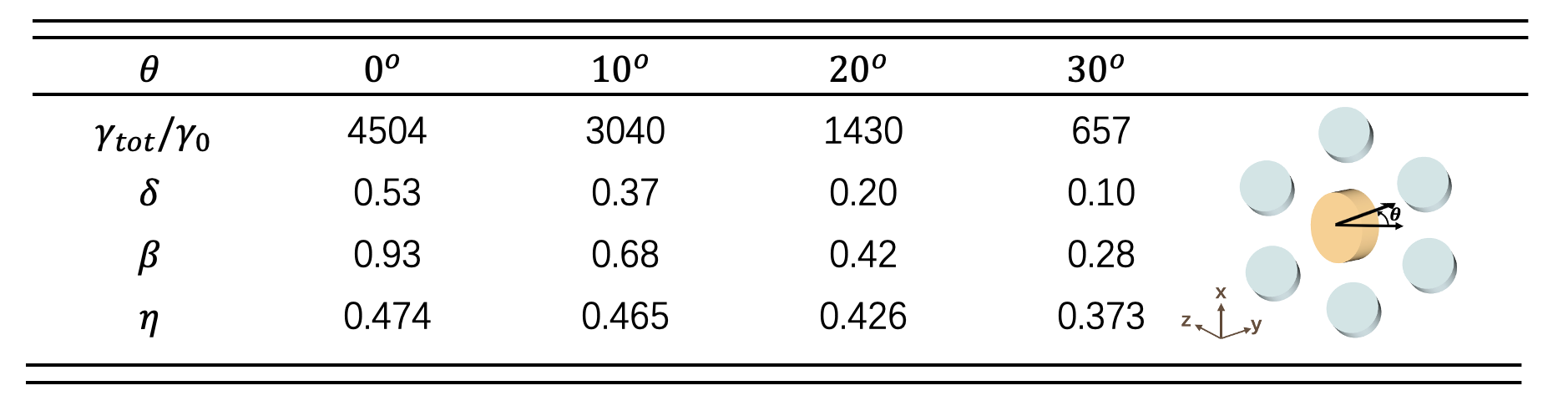}% Here is how to import EPS art
\end{table*}

%\section{collection}\label{Sec2}
For practical application of single photon sources, besides large emission rate, efficient collection of photons is also extremely important.
For a nanodisk in free space, emitted photons are difficult to be collected and utilized.
In the topological hybrid structure, the topological PC can guide the photons that were previously scattered to the surroundings into the edge states. If the collection efficiency is defined as $\beta = \gamma_{ed}/\gamma_{tot}$, $\beta$ can reach 93\% [inset in Fig. 2(c)].
The high collection efficiency in the hybrid structure comes from the large near-field overlapping between equivalent large magnetic dipole and the edge state as well as anti-scattering of topological state  [Fig. 2(b)].
In previously reported enhanced emission via hybrid structures containing metallic nanoantenna~\cite{S8,S9,S10}, quantum yield is not very high. 
But here, owing to no absorption loss in dielectric structures and anti-scattering property of edge states, the quantum yield is almost equal to the collection efficiency , i.e., almost all the emitted photons can be used for on-chip photonic devices. It is worth noting that the narrow linewidth in the hybrid structure also comes from the fact that edge states can collect scattered photons~\cite{jia}.

%\section{overlapping degree}\label{Sec2}

%The emission cascade enhancement and efficient photon collection obtained in the topological hybrid structure are due to the large near-field overlapping between the resonant nanodisk and the edge state.

To quantitively describe the near-field overlapping between the magnetic dipole resonance of the nanodisk and topological edge state,  we define the overlapping  degree  as $\eta =\int \frac{\left| \vec{E}_{1} \cdot  \vec{E}_{2} \right|}{E{}^{2}_{0}V_{m}} dV$, where $\vec{E}_{1}$, $\vec{E}_{2}$, and $\vec{E}_{0}$ are the electric fields of the edge state,  the resonant nanodisk without the PC,   and  the system background, respectively, and $V_m$ is the calculation area~\cite{SM}.
By rotating the nanodisk around the x-axis by an angle $\theta$ [inset in Table I], it can be seen that as $\theta$ increases, their near-field overlapping degree $\eta$  decreases [Table I].
When $\theta$ becomes larger, the polarization of the equivalent large magnetic dipole is shifted, then the number of photons to excite the edge state is smaller, i.e., their near-fields  less overlap. As a result, there is a decrease of  the cascade enhancement factor $\delta$ as well as magnetic Purcell factor  $\gamma_{tot}/\gamma_{0}$.
 Meanwhile, the efficiency of collecting photons along the edge states also becomes lower.

%\section{disuss}\label{Sec2}

%We now discuss the effects of the structural parameters of nanodisks on the cascade enhancement and collection efficiency of single photon emission.
%When the dielectric constant $\varepsilon_n=16$ and radius $R_n=$75 nm of the nanodisk are fixed, it can be seen that with the increase of height $H_n$, the magnetic Purcell factors $\gamma_{tot}/\gamma_{0}$ and the cascade enhancement factor $\delta$ both increase[Fig. 3(a)].
%Similarly, when fixing the dielectric constant  $\varepsilon_n=16$ and the height $H_n$=89 nm of the nanodisk, the magnetic Purcell factors and the cascade enhancement factor increase with the increase of the radius $R_n$ [Fig. 3(b)].
%This is because larger nanodisks have more near-field overlap with edge states [inset in Fig. 3].
%It should be noted that the collection efficiency is insensitive to changes in the size of the nanodisk. The collection efficiency is almost constant when changing the structural parameters of the nanodisk~\cite{SM}.
%This is mainly because even a relatively small nanodisk can be equivalent to a large $y$-polarized magnetic dipole, thereby fully exciting the edge state.
We now discuss the effects of the structural parameters of nanodisks on the cascade enhancement and collection efficiency of single photon emission.
%We performed a systematic study by varying the radius $R_n$ and height $H_n$ of the nanodisk separately.
As the height $H_n$ increases, both the Purcell factor and the cascade enhancement factor become larger [Fig. 3(a)].
Then, we find that the expansion of the radius $R_n$ also leads to higher Purcell factor and cascade enhancement factor [Fig. 3(b)]. 
The reason is that larger nanodisks have more near-field overlapping with edge states [inset in Fig. 3].
In addition, when the size of the nanodisk is enlarged, the wavelength of magnetic dipole resonance redshifts, so the spontaneous emission spectrum in the hybrid structure also redshifts.
It should be noted that the collection efficiency  is insensitive to the size of the nanodisk, remaining almost constant.~\cite{SM}.

\begin{figure*}[htbp]
\includegraphics[width=1\textwidth]{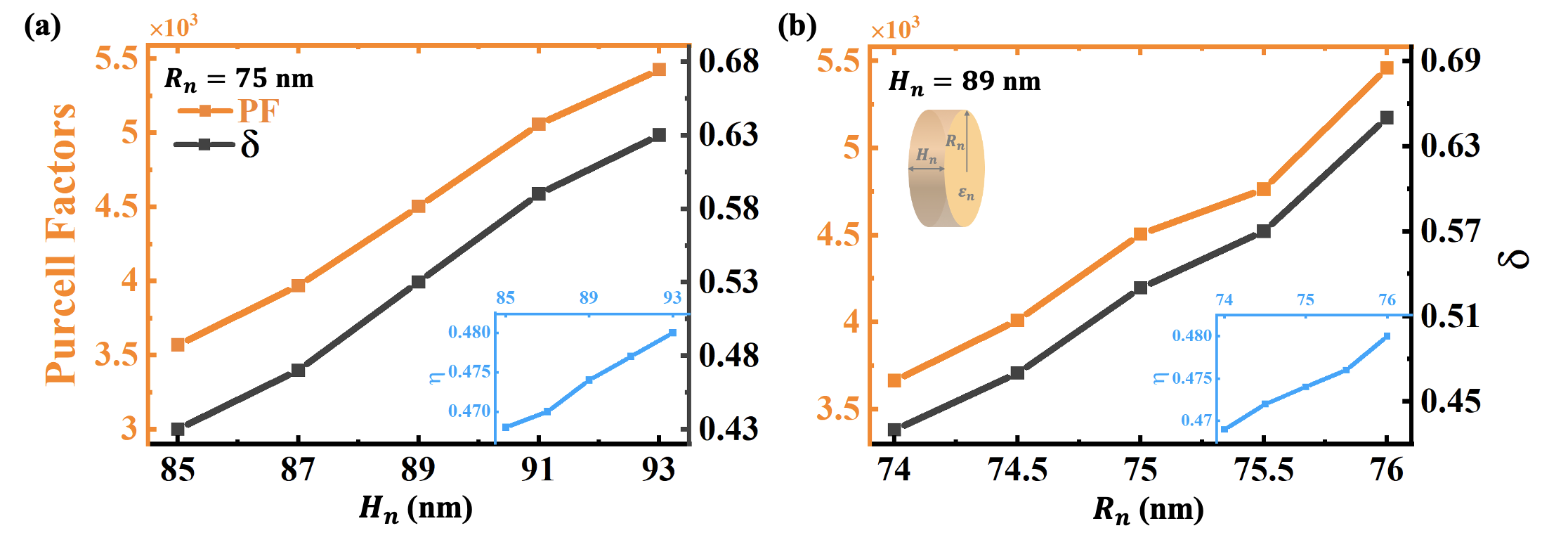}% Here is how to import EPS art
\caption{The influence of nanodisk's size on the response of topological hybrid structure. Purcell factor, cascade enhancement factor, and near-field overlapping (a) with different heights $H_n$ of nanodisks, (b) with different radius $R_n$ of nanodisks. As the size of the nanodisk increases, the near-field overlapping increases, so the Purcell factor and the cascade enhancement factor increase. Here, the dielectric constant is $\varepsilon_n=16$.}\label{fig4}
\end{figure*}

%\section{disuss}\label{Sec2}
In the topological hybrid structure, as long as the y-polarized magnetic quantum emitter is placed in the near-field region of the nanodisk, it can excite the magnetic dipole resonance, thereby obtaining a cascade enhancement of the Purcell factor.
Since the magnetic hotspot is at the center of the disk, the maximum Purcell factor can be obtained when the quantum emitter is placed at the center, compared with the cases far away from the hot spot~\cite{SM}. In addition, even if the position of the nanodisks in the topological PC is slightly shifted, the emission cascade enhancement phenomenon can still be observed~\cite{SM}, which facilitate the experimental realization.

The magnetic emission cascade enhancement is sensitive to the polarization of the quantum emitter. Only when the magnetic dipole resonance of the nanodisk is sufficiently excited, does the large near-field overlapping with the edge states occur, resulting in a cascade enhancement of the emission. Since the x-polarized and z-polarized dipoles excite the magnetic dipole resonance of the nanodisk insufficiently, there is no large emission enhancement~\cite{SM}. 
This mechanism of emission enhancement has the potential application in on-chip single-photon sources.
%so stability is also very important. 
%We theoretically demonstrate that even if there are defects and perturbations in the topological hybrid structure, cascade enhancement of the emission and efficient collection efficiency can still be achieved~\cite{SM}. This topological protection is beneficial for experimental implementation.
When the structural defect is created by removing the cylinder from the PC, the cascade enhancement factor and collection efficiency are almost the same as those without the defect~\cite{SM}. This topological protection is beneficial for experimental implementation.

%\section{experimentally}\label{Sec2}
Finally, we discuss the possibility of experimental implementation of our scheme.
At present, both topological PC~\cite{M4} and dielectric nanocavities~\cite{M5} can be manufactured using nanotechnology, e.g., electron beam lithography and electrochemical etching. Single emitters can be realized in many forms, such as molecules
~\cite{M6}, Rydberg atoms~\cite{M7}, and quantum dots~\cite{M8}. In some quantum emitters such as rare earth ions~\cite{M9} and semiconductor quantum dots~\cite{M10}, there are significant magnetic dipole transitions whose intensities are comparable to or even stronger than the competing electric dipole transitions. 
%The integration of semiconductor quantum dots into topological waveguides has also been achieved~\cite{M11}. Therefore, we propose that the mechanism of cascade enhancement of emission and efficient collection of photons under topological protection may be experimentally realized in the future.
It is possible to place emitters into the hybrid topological structure, such as by integrating semiconductor quantum dots into topological waveguides ~\cite{M11} or doping rare-earth ions into nanodisk~\cite{M12}. Therefore, the scheme we proposed here may be experimentally realized in the future.

%\section{summary}\label{Sec2}
\section{Conclusion}
In summary, we have proposed a mechanism for cascade enhancement and efficient collection of emitted photons under topological protection.
%In the hybrid structure of topological PC containing a resonant nanodisk, the magnetic dipole resonance of the nanodisk can be equivalent to a large magnetic dipole to interact with the edge state.
%The large near-field overlapping between the resonant nanodisk and the edge state enables cascade enhancement of magnetic emission.
%At the same time, photons that were previously scattered around can be efficiently guided to propagate in the edge state channel, the collection efficiency can reach more than 90\% due to the robustness of the edge state.
%The low loss of the dielectric structure and the anti-scattering properties of the topological mode make the collection efficiency almost equivalent to the quantum yield, that is, almost all of the collected photons can be used in on-chip photonic devices.
%This mechanism will provide new ideas for the study of cavity quantum electrodynamics in topological structures.
%A hybrid structure integrating nanodisk and topological PC is designed, in which leverages the nanodisk's magnetic dipole resonance to interact strongly with edge states.
A hybrid structure integrating topological PC and nanodisk is designed, in which the magnetic dipole resonance of the nanodisk can be equivalent to a large magnetic dipole to interact with the edge state.
The large near-field overlapping between the resonant nanodisk and the edge state enhances magnetic emission through a cascade effect.
At the same time, emitted photons can be efficiently guided to propagate in the edge state channel, achieving a collection efficiency exceeding 90\% due to the robustness of the edge state.
The low loss of the dielectric structure and the anti-scattering properties of the topological mode make the quantum yield almost equivalent to the collection efficiency, that is, almost all of the emitted photons can be used in on-chip photonic devices.
The mechanism proposed here enhances the spontaneous emission rate of the emitter while achieving large collection efficiency under topological protection, which will provide practical use for ultra-bright and stable on-chip single photon sources. It also provides new insights for the study of cavity quantum electrodynamics in topological structures.

%%%%%%%%%%%%%%%%%%%%%%%%%%%%%%%%%%%%%%%%%%%%%%%%%%%%%%%%%%%%%%%%%%%%%
%% The "Acknowledgement" section can be given in all manuscript
%% classes.  This should be given within the "acknowledgement"
%% environment, which will make the correct section or running title.
%%%%%%%%%%%%%%%%%%%%%%%%%%%%%%%%%%%%%%%%%%%%%%%%%%%%%%%%%%%%%%%%%%%%%
\begin{acknowledgement}
This work is supported by the National Natural Science Foundation of China under Grants No. 11974032 and the Innovation Program for Quantum Science and Technology under Grant No. 2021ZD0301500.

\end{acknowledgement}

%%%%%%%%%%%%%%%%%%%%%%%%%%%%%%%%%%%%%%%%%%%%%%%%%%%%%%%%%%%%%%%%%%%%%
%% The same is true for Supporting Information, which should use the
%% suppinfo environment.
%%%%%%%%%%%%%%%%%%%%%%%%%%%%%%%%%%%%%%%%%%%%%%%%%%%%%%%%%%%%%%%%%%%%%
\begin{suppinfo}

%A listing of the contents of each file supplied as Supporting Information
%should be included. For instructions on what should be included in the
%Supporting Information as well as how to prepare this material for
%publications, refer to the journal's Instructions for Authors.
%
%The following files are available free of charge.
%\begin{itemize}
%  \item Filename: brief description
%  \item Filename: brief description
%\end{itemize}

\end{suppinfo}

%%%%%%%%%%%%%%%%%%%%%%%%%%%%%%%%%%%%%%%%%%%%%%%%%%%%%%%%%%%%%%%%%%%%%
%% The appropriate \bibliography command should be placed here.
%% Notice that the class file automatically sets \bibliographystyle
%% and also names the section correctly.
%%%%%%%%%%%%%%%%%%%%%%%%%%%%%%%%%%%%%%%%%%%%%%%%%%%%%%%%%%%%%%%%%%%%%

\newpage
\section{Graphical TOC Entry}
\begin{figure*}[htbp]
\includegraphics[width=8.25cm]{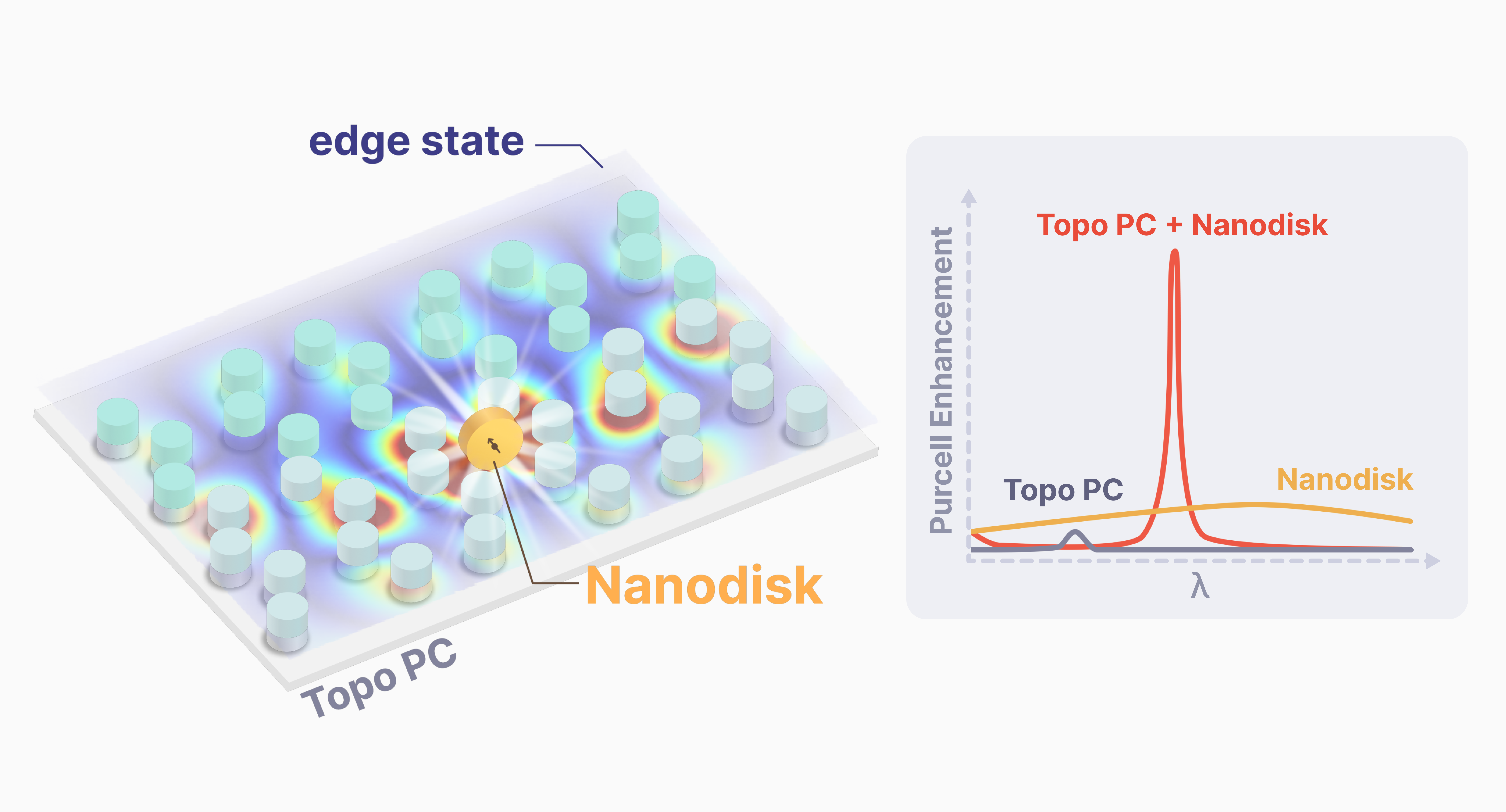}% Here is how to import EPS art
\end{figure*}
\end{document}